\documentclass[preprint,aps,showpacs]{revtex4}
\usepackage{graphicx,amsfonts}
\usepackage{epsfig}
\usepackage{dcolumn}
\usepackage{bm}
\begin{document}

\title{Grand Unification and Proton Stability \\ 
Near the Peccei-Quinn Scale}

\author{Alex G. Dias\footnote{e-mail: alexdias@fma.if.usp.br}}
\affiliation{Instituto de F\'\i sica, Universidade de S\~ao Paulo, \\
C. P. 66.318, 05315-970 S\~ao Paulo, SP, Brazil. }

\author{V. Pleitez\footnote{e-mail: vicente@ift.unesp.br}}
\affiliation{Instituto de F\'\i sica, Te\'orica, Universidade Estadual
Paulista \\
Rua Pamplona 145, S\~ao Paulo, SP, Brazil. }

\begin{abstract}
We show that in an $SU(2)\otimes U(1)$ model with a DSF-like invisible axion it
is possible to obtain (i)  the convergence of the three gauge coupling constants
at an energy scale near the Peccei-Quinn scale; (ii) the correct value for 
$\sin^2\hat{\theta}_W(M_Z)$; (iii) the stabilization of the proton by the cyclic
$Z_{13}\otimes Z_3$ symmetries which also stabilize the axion as a solution to
the strong $CP$ problem. Concerning
the convergence of the three coupling constants and the prediction of the weak
mixing angle at the $Z$-peak, this model is as good as the minimal
supersymmetric standard model with $\mu_{\rm SUSY}=M_Z$. We also consider the
standard model with six and seven Higgs doublets.  The main calculations were
done in the 1-loop approximation but we briefly consider the 2-loop
contributions. 

\end{abstract}
\maketitle

The convergence of the three gauge coupling constants
$g_3,g_2,g_1=\sqrt{5/3}g^\prime$, and the prediction of the electroweak mixing
angle are some of the motivations for grand unified theories
(GUTs)~\cite{gg74,hquinn,gut,gut2}. Unfortunately the simplest and more elegant
GUTs which break into a simple step to the standard model, like
$SU(5)$~\cite{gg74,hquinn} and some of the $SO(10)$ and $E_6$
GUTs~\cite{gut,gut2}, were ruled out by two experimental results. The first one
is concerned with the fact that using the value of the electroweak mixing angle
measured at the $Z$-peak by LEP, \textit{i.e.},
$\sin^2\hat{\theta}_W(M_Z)=0.23113(15)$~\cite{pdg}, the running coupling
constants extrapolated from the values measured at low energies do not meet at a
single point or, inversely, assuming the convergence of the three coupling
constants, the prediction of the value of $\sin^2\hat{\theta}_W(M_Z)$ (or,
alternatively, we can obtain the convergence point of $\alpha_1$ and $\alpha_2$
and then the value of $\alpha_s(M_Z)$ can be predicted) does 
not agree with the experimental value. On the other hand, in the minimal
supersymmetric extension of the standard model (MSSM for short) the coupling
constants do intercept at a single point, with a remarkable precision, and the
predicted value of the weak mixing angle is also in agreement with the observed
value~\cite{susy}. In fact, this has become one of the most important reason for
believing on the existence of supersymmetry at the TeV scale and, in particular,
on the MSSM. In particular, a new scale  for physics, $\mu_{\rm SUSY}$, is then
required. The second trouble with the simplest GUTs  is related with the
non-observation of the proton decay at the predicted lifetime. Recent
data by Super-Kamiokande on $p\to K^+\bar{\nu}$ imply that $\tau_P>10^{33}$
years~\cite{pdg,sk99}. However, this is also a trouble for SUSY $SU(5)$ since
this theory has $d=5$ effective operators induced by colored-Higgs triplet
that produce a rapid proton decay~\cite{d5} and it is necessary to
appeal to fermion mixing in order to keep (tight) agreement with
data~\cite{perez}. 
Thus, it appears natural to ask ourselves if there are options for SUSY 
yielding the convergence of the couplings, the observed value of the weak mixing
angle at the $Z$-pole and an appropriately stable proton. The importance of the
Higgs boson contributions to the convergence of the couplings has been
emphasized recently~\cite{willenbrock}, although exotic fermions can also
lead to the gauge coupling unification even at the TeV
range~\cite{berezhiani01}.  
In this cases, unification at lower energy scale is possible even without
supersymmetry and the proton stability is guaranteed by
additional assumptions as extra dimensions or dynamical symmetry  
breaking~\cite{berezhiani01}, by the conservation of the baryon number in the
gauge interactions as in the $[SU(3)]^3$ trinification~\cite{willenbrock}, or in
$[SU(3)]^4$ quartification where the proton decay is mediated only by Higgs
scalars~\cite{bmw}.   

Here, we will show that in a  multi-Higgs extension of the standard
model we have the convergence of the three gauge  coupling constants at an
energy scales of the order of $10^{13}$ GeV, and the weak 
mixing angle in agreement with the measured value at the $Z$-pole. The proton
decay occurs only throughout dimension 8,9,10 and 11 effective operators because
discrete $Z_{13}\otimes Z_3$ symmetries forbid all $d=6,7$ operators.  
This model was proposed independently of the issue of the gauge unification, and
it has a DFS-like invisible axion~\cite{dine} stabilized against semiclassical
gravitational effect by those discrete symmetries~\cite{iaxionsm}. The use
of discrete gauge symmetries in the proton decay problem has been used recently
in Ref.~\cite{gogoladze} in a model with a $Z_6$ symmetry, which baryon number
violation  low dimension dangerous effective operators are all forbidden. 
For other recent uses of symmetries like these see~\cite{outros}.
We compare our results with the usual MSSM showing that the convergence of the
coupling constants and the prediction of the $\sin^2\hat{\theta}_W(M_Z)$ is, in
this model, as good as in the MSSM with $\mu_{\rm SUSY}=M_Z$. 
We also consider the SM with seven Higgs doublets, and briefly comment 
the SM with six Higgs doublets and $\mu_{\rm SUSY}=1$ TeV. 

In Ref.~\cite{iaxionsm} the representation content of the standard model was
augmented, by adding scalar fields and right-handed neutrinos, in such a way
that the discrete $Z_{13}\otimes Z_3$ symmetries could be implemented in the
model.  
The particle content of the model is the following:  
$Q_L=(u\,d)^T_L\sim({\bf2},1/3)$, $L_L=(\nu\,l)^T_L\sim({\bf2},-1)$ denote any
quark and lepton doublet; $u_R\sim({\bf1}, 4/3)$, $d_R\sim({\bf1},-2/3)$,
$l_R\sim({\bf1},-2)$, $\nu_R\sim({\bf1},0)$ are the respective right-handed
components. It was also assumed that each charge sector gain mass from a
different scalar doublet, hence we have the following Higgs multiplets: four
doublets $\Phi_u$,  $\Phi_d$, $\Phi_l$ and $\Phi_\nu$ [all of them of the
form $({\bf2},+1)=(\varphi^+,\,\varphi^0)^T$ under $SU(2)\otimes U(1)$] which
generate Dirac masses for $u$- and $d$-like quarks, charged leptons and
neutrinos, respectively; a neutral complex singlet $\phi\sim({\bf1},0)$, a
singly charged singlet $h^+\sim({\bf1},+2)$ and, finally, 
a non-hermitian triplet $\vec{{\cal T}}\sim({\bf3},+2)$.  With the discrete 
symmetries flavor changing neutral currents are also naturally suppressed at the
tree level.  

Next, let us consider the evolution equations of the three gauge coupling
constants, at the 1-loop level,
\begin{eqnarray}
\alpha^{-1}_1(M)&=&\alpha^{-1}(M_Z)\,\frac{3}{5}\cos^2\theta_W(M_Z)+
\frac{b_1}{2\pi}\,\ln\left(\frac{M_Z}{M}
\right),\nonumber \\ 
\alpha^{-1}_2(M)&=&\alpha^{-1}(M_Z)\sin^2\theta_W(M_Z)+
\frac{b_2}{2\pi}\,\ln\left(\frac{M_Z}{M} \right),\nonumber \\
\alpha^{-1}_3(M)&=&\alpha^{-1}_3(M_Z)+
\frac{b_3}{2\pi}\,\ln\left(\frac{M_Z}{M} \right), 
\label{evolution}
\end{eqnarray}
where $b_i$ are the 1-loop beta-function coefficients
\begin{equation}
b_i=\frac{2}{3}\sum_{\rm fermions}T_{Ri}(F)+
\frac{1}{3}\sum_{\rm scalars}T_{Ri}(S)-\frac{11}{3}C_2(V).
\label{bi}
\end{equation}
For $SU(N)$, we have that $T_R=C_2=N$,  with $N\geq 2$, for fields in the 
adjoint representation, and $T_R(S,F)=1/2$ for fields in the fundamental or
antifundamental representation; for $U(1)$, $C_2(V)=0$ and $T(S_a,F_a)=(3/5){\rm
Tr}(Y^2_a/4)$ for a unification in $SU(5)$ [it is the same for the case of
$SO(10)$]. Eq.~(\ref{bi}) is valid for Weyl spinors and complex scalar fields. 
Considering the extension of the standard model 
with $N_g$ matter generations, $N_H$ Higgs doublets and $N_T$ non-hermitian
scalar triplets, all of them considered relatively light, 
\begin{eqnarray}
& &b_1=\frac{4}{3}N_g+\frac{1}{10}N_H+\frac{3}{5}N_T,\nonumber \\
& &b_2=\frac{4}{3}N_g+\frac{1}{6}N_H+\frac{2}{3}N_T-\frac{22}{3},\nonumber \\
& &b_3=\frac{4}{3}N_g-11.
\label{bi2}	
\end{eqnarray}
Only the scalar singlet $h^+$ will be considered with mass of the order of the
unification scale. 
The evolution equations in Eq.~(\ref{evolution})
implies the unification condition 
$\alpha^{-1}_1(M_U)=\alpha^{ 1}_2(M_U)=\alpha^{-1}_3(M_U)\equiv \alpha^{-1}_U$, 
which also defines the mass scale $M_U$:
\begin{equation}
M_U= M_Z\exp\left[{2\pi\,\frac{{\alpha^{-1}(M_Z)}-
\frac{8}{3}{\alpha_3^{-1}(M_Z)}}{ 
\frac{5}{3}b_1+b_2-\frac{8}{3}b_3
}}\right]. 
\label{rge}
\end{equation}

\begin{figure}[ht] 
\begin{center} 
\leavevmode 
\mbox{\epsfig{file=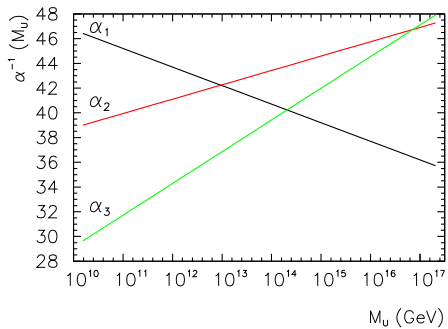,width=0.4\textwidth,angle=0}}        
\end{center} 
\caption{Running couplings in the standard model.} 
\label{321} 
\end{figure}

From Eqs.~(\ref{bi2}) with $N_g=3$, $N_H=1$ and $N_T=0$, \textit{i.e.}, the
standard model, give $(b_1,b_2,b_3)=(41/10,-19/6,-7)$. In this case, the
evolution of $g_i$ is shown in Fig.~\ref{321}.
In this figure (and below) we have used the inputs $M_Z=91.1876$ GeV;
$\alpha(M_Z)=1/128$; and $\alpha_3(M_Z)=0.1172$ \cite{pdg}. It is clear that
with only the representation content of the SM, there is no convergence of the
three $g_i$ at a given point \cite{susy}. 

\begin{figure}[ht] 
\begin{center} 
\leavevmode 
\mbox{\epsfig{file=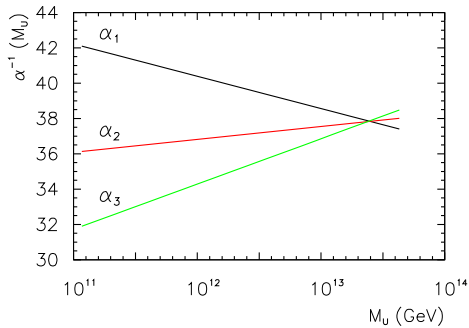,width=0.4
\textwidth,angle=0}}        
\end{center} 
\caption{Running couplings in the present model.}
\label{321ex} 
\end{figure}

On the other hand, with $N_g=3,N_H=4$
and $N_T=1$ \textit{i.e.},  the model of Ref.~\cite{iaxionsm}, with
Eqs.~(\ref{bi2}) we have $(b_1,b_2,b_3)=(5,-2,-7)$ and the evolution of the
coupling constants in this case is shown in Fig.~\ref{321ex}.  
We obtain that  the three forces unify at the energy scale 
$M_U\simeq 2.8\times 10^{13}$ GeV and $\alpha^{-1}_U\simeq 38$. 
The value predicted for $\sin^2\hat{\theta}_W(M_Z)$, using the unification scale
as an input, is

\begin{equation}
\sin^2\hat{\theta}_W=\frac{3}{8}+\frac{5}{16\pi}\alpha(M_Z)(b_1-b_2)\,\ln\left(
\frac{M_Z}{M_U}\right),
\label{sinz}
\end{equation}
and we obtain in this model $\sin^2\hat{\theta}_W(M_Z)=0.2311$, which coincides
with the measured value  as it should be since the unification occurs with a
very good precision. In order to compare this result with
others~\cite{willenbrock} we write at the 1-loop approximation, by
eliminating the unification scale from Eq.~(\ref{bi}),   
\begin{eqnarray}
\tilde{b}&=& \frac{11+\frac{1}{2}N_H+2N_T}{22-\frac{1}{5}(N_H+N_T)} 
\nonumber \\
&=&\frac{5}{3}\,\left[\frac{\sin^2\hat{\theta}_W(M_Z)-\frac{\alpha(M_Z)}
{\alpha_3(M_Z)}}{1-\frac{8}{3}\sin^2\hat{\theta}_W(M_Z)}\right],
\label{btilde}
\end{eqnarray}
where $\tilde{b}=(b_3-b_2)/(b_2-b_1)$~\cite{berezhiani01}. 
For being more general we let $N_H$ and $N_T$ arbitrary to see 
if there are other values for them which could fit with  unification. 
The theoretical ratio $\tilde{b}$ defined in the first line of
Eq.~(\ref{btilde}), and which depends mainly on the scalar representation
content (notice that, at the 1-loop level, $\tilde{b}$ does not depend on
$N_g$), should coincide with the quantity defined in the second line which
depends only on  the experimental values of the coupling constants $\alpha$ and
$\alpha_3$ and $\sin^2\hat{\theta}_W$ at the $Z$-peak. The experimental inputs
then implies $\tilde{b}=0.714$ using the second line of Eq.~(\ref{btilde}). 
Using the first line of Eq.~(\ref{btilde}) the minimal standard model 
implies $\tilde{b}=115/218\simeq 0.527$ (including the scalar contributions), 
so that it does not match and $\sin^2\hat{\theta}_W(M_Z)=0.204$ according to
Eq.~(\ref{sinz}) and for  this reason the unification of this model 
in non-supersymmetric $SU(5)$ was ruled out by LEP data. In the present model we
obtain $\tilde{b}=5/7\simeq 0.714$ and this value matches in Eq.~(\ref{btilde})
and gives, then, the observed value for $\sin^2\hat{\theta}_W(M_Z)$~\cite{pdg} 
as we pointed out above.

\begin{figure}[ht] 
\begin{center} 
\leavevmode 
\mbox{\epsfig{file=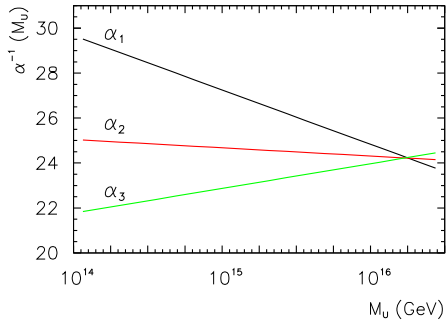,width=0.4
\textwidth,angle=0}}        
\end{center} 
\caption{Running couplings in the MSSM with $\mu_{\rm SUSY}=M_Z$.} 
\label{mssmmz} 
\end{figure}

In the MSSM when $\mu_{\rm SUSY}=M_Z$ we have $(b_1,b_2,b_3)=(33/5,1,-3)$ and
the respective evolution is shown in Fig.~\ref{mssmmz}, with $M_U\simeq
2.1\times10^{16}$ GeV and the inverse of the coupling constant at the
unification scale, denoted in this case by $\alpha_5$, is $\alpha^{-1}_5\simeq
24$. The weak mixing angle has also the correct value~\cite{marciano82} and
the same value for $\tilde{b}$ is obtained like in the present model. The case
when the SUSY scale is of the order of $M_Z$ is better than the case when that
scale is of the order of 1 TeV but we do not show the latter case. As can be
seen from Figs. 2 and 3 and from the value of $\tilde{b}$, concerning the
unification and and the prediction of the weak mixing angle, the model of
Ref.~\cite{iaxionsm} is as good as the MSSM and for this reason it was not
necessary in the present work to take into account of the theoretical and
experimental uncertainties.

\begin{figure}[ht] 
\begin{center} 
\leavevmode 
\mbox{\epsfig{file=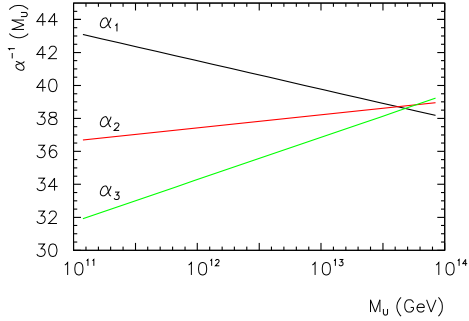,width=0.4
\textwidth,angle=0}}        
\end{center} 
\caption{Case of the standard model with seven scalar doublets.} 
\label{3217d} 
\end{figure}

Finally, in the case when $N_g=3$ but $N_H=7$ and $N_T=0$ we have 
$(b_1,b_2,b_3)=(47/10,-13/6,-7)$, $\tilde{b}= 145/206\simeq 0.704$ and 
$\sin^2\hat{\theta}_W(M_Z)=0.230$ according to Eq.~(\ref{sinz}) and the
evolution is shown in Fig.~\ref{3217d} with $M_U\simeq 5.8\times10^{13}$
GeV~\cite{gut}. We have also studied the case of the SM with six Higgs doublets.
In this case the unification of the coupling coincides with that of
Ref.~\cite{willenbrock} and $\sin^2\hat{\theta}_W(M_Z)=0.226$. Moreover, the
convergence of the three coupling constants in the seven Higgs scalar doublets
is better than the case of six of such doublets~\cite{willenbrock}.  

We see that, in order to have the unification of the couplings and the correct
value for the weak mixing angle at the $Z$-pole it is not necessary 
to have low energy supersymmetry. On the other hand, four Higgs
doublets and a non-hermitian Higgs triplet are more precise, for the unification
of the coupling constant, than just seven Higgs doublets. As we said before,
only the singlet $h^+$ has been considered having a mass of 
the order of the unification scale. Assuming that this singlet is light we
obtain the same value for $M_U$ but $\sin^2\hat{\theta}_W(M_Z)=0.229$. 

The contribution of just one Higgs doublet is almost 
negligible compared to that of quarks and leptons when dealing with the
renormalization group  equations. But in models like the multi-Higgs extensions
we have discussed here, the Higgs multiplets have a total contribution already
important to unify the theories. However, we stress again, looking at
the Eq.~(\ref{btilde}), that the value of $\sin^2\hat{\theta}_W(M_Z)$ depends
strongly on $N_{H}$ and $N_{T}$ selecting in this way only few possibilities. 
Moreover, the prediction of the weak mixing angle in this model is not an
accident. To see that, with the experimental inputs in the second line of
Eq.~(\ref{btilde}) we have  $N_H=7.324-3.333N_T$
and it is clear that the best solution for $N_H$ and $N_T$ integers is when
$N_H=4$  for $N_T=1$. The seven doublets model $N_H=7$ for $N_T=0$ is the second
best solution. The denominator in the exponent of Eq.~(\ref{rge}) can be written
as $(1/3)N_H+(5/3)N_T+22$, and we see that a larger number of Higgs multiplets
imply a smaller value for $M_U$, however, we obtain the best match condition for
$\tilde{b}$   with $N_H=4$ and $N_T=1$. 

Due to the precise measurements in the electroweak sector higher order
corrections to the 1-loop calculations are important to verify if the solutions
above can be stable against these corrections. Considering only the gauge
coupling constants, their running $g_i(\mu)$ are now solutions of the corrected
renormalization group equations
\begin{equation}
\mu\frac{d\, \alpha_i(\mu)}{d\, \mu}= \frac{1}{2\pi}  
\left [ b_i+\frac{1}{4\pi}\sum_{j=1}^{3} b_{ij}
\alpha_j(\mu)\right ]\alpha_i(\mu)^2.
\label{rg2loop}
\end{equation}
The general form of the coefficients $b_{ij}$ is given in
Refs.~\cite{jones}. For the cases discussed in this paper, \textit{i.e.}, the only
nontrivial scalar representations under $SU(2)$ are doublets and triplets, they have the
following form
\begin{eqnarray}
b_{ij}= \left(\begin{array}{ccc}
  \frac{19}{15}N_g+\frac{9}{50}N_H+
  \frac{36}{5}N_T & \frac{3}{5}N_g+
  \frac{9}{10}N_H+\frac{72}{5}N_T & \frac{44}{15}N_g \\
  \frac{1}{5}N_g+\frac{3}{10}N_H+\frac{24}{5}N_T & \frac{49}{3}N_g
  +\frac{13}{3}N_H+\frac{56}{3}N_T-\frac{136}{3}& 4N_g\\
  \frac{11}{30}N_g & \frac{3}{2}N_g & \frac{76}{3}N_g -102
\end{array}\right). 
\label{bij}
\end{eqnarray}

The quark top Yukawa coupling being of order of unity, is the only one
comparable with the $\alpha_i$. However its contribution to
Eqs.~(\ref{rg2loop}), are  unimportant when compared with the other
contributions in Eq.~(\ref{bij}), and does not affect the 2-loops running
significantly for the  values of $N_H$ and $N_T$ considered here. 
A complete treatment includes, of course, all scalar interactions (which in the
present model include trilinear interactions) and it is much more complicated.
However, as an illustration, we will consider only the corrections of the gauge
coupling constants in Eq.~(\ref{bij}). In this case, the numerical solutions to
the system of equations in Eqs.~(\ref{rg2loop}) can be found using the
Eq.~(\ref{evolution}) after making the simply substitution $b_i\rightarrow
\Delta_i$, with the $\Delta_i$ extracted numerically. The values of $\Delta_i$
and the respective values for $M_U$ and $\sin^2\hat{\theta}_W(M_Z)$, at the 
2-loop order, are given in Table~\ref{table1}. We see from the table that in the
cases with $N_H=4$, $N_T=1$ and $N_H=7$, $N_H=0$, the values of the weak mixing
angle at the $Z$ pole are a little above the experimental value (since this
is only an illustration that consider just the evolution of the gauge coupling
constants it is not necessary to take into account the experimental and
theoretical uncertainties). 
This does not rule out these models since the correct value of 
$\sin^2\hat{\theta}_W(M_Z)$ may be obtained at the 2-loop level by constraining 
the self-interactions couplings and the masses of the scalars fields. Hence,
this partial analysis suggests that solutions in the 1-loop approximation might
be stable under higher corrections since they are not drastically changed when
2-loop corrections are included.   
\begin{table}
\begin{tabular}{||c|c|c|c|c|c|c||}\hline
$N_H$ & $N_T$ & $\Delta_1$ & $\Delta_2$ & $\Delta_3$ & $M_U$ & 
$\sin^2\hat{\theta}_W(M_Z)$\\
\hline
4 & 1 & 5.102 & -1.847 & -7.089 & 1.5 & 0.235\\ 
6 & 0 & 4.659 & -2.213 & -7.089 & 5.0 & 0.231\\
7 & 0 & 4.762 & -2.035 & -7.089 & 3.4 & 0.234\\ \hline
\end{tabular}
\caption{Values for $\Delta_i$  and 
$\sin^2\hat{\theta}_W(M_Z)$ at the 2-loop level for multi-Higgs models.
$M_U$ is in units of $10^{13}$ GeV.}
\label{table1}
\end{table}

Next, we come to the question of the proton stability in the present model. 
We have seen that the energy scale of the unification of the coupling constants
is of the order of $10^{13}$ GeV, \textit{i.e.}, smaller than the scale of the
non-SUSY $SU(5)$ and near, by a 10-1000 factor, to the Peccei-Quinn scale. This
is, apparently, a disaster from the point of view of the nucleon decay. However,
it is not because the model accept discrete symmetries that forbid potentially
dangerous effective operators of $d=6,7$. With the representation content of the
model it is possible to impose the following $Z_{13}$ symmetry:   
\begin{eqnarray}
& &Q\to \omega_5Q,\; u_R \to \omega_3u_R,\; d_R\to \omega^{-1}_5 d_R,
\;L\to \omega_6 L,
\nonumber \\
& & \nu_R\to \omega_0\nu_R, \;l_R\to \omega_4 l_R,  
\Phi_u\to \omega^{-1}_2\Phi_u,\; \Phi_d\to \omega^{-1}_3\Phi_d, \nonumber \\
& &\Phi_l\to \omega_2\Phi_l,\;\Phi_\nu\to \omega^{-1}_6\Phi_\nu, 
\phi\to\omega^{-1}_1\phi,\; T\to w^{-1}_4T,\nonumber \\
& &h^+\to \omega_1h^+,
\label{z13}
\end{eqnarray}
with $\omega_k=e^{2\pi ik/13},\;k=0,1,...,6$. Moreover, in order to
have an automatic PQ symmetry~\cite{pq}, it is also necessary to impose a $Z_3$
with parameters denoted by $\tilde{\omega}_0$, $\tilde{\omega}_1$, and
$\tilde{\omega}^{-1}_1$ with $\Phi_l$ transforming with $\tilde{\omega}^{-1}_1$;
$\Phi_\nu$, $\nu_R$, $l_R$ with $\tilde{\omega}_1$, while all other fields
transform trivially under $Z_3$. For details see Ref.~\cite{iaxionsm}.

We search for effective operators that are $SU(3)_C\otimes SU(2)_L\otimes
U(1)_Y$ and $Z_{13}\otimes Z_3$ invariant. No grand unified model is assumed
here. Effective operators with $d=6$ which induce a rapid proton
decay for $M_U<10^{16}$ GeV, are given by~\cite{sw79}:  
\begin{eqnarray}
&&O^{(1)}_{abcd}=\epsilon_{ijk}\epsilon_{\alpha\beta}\overline{(d_R)^c_{ia}}
(u_{R})_{jb}\overline{(Q_L)^c_{k\alpha c}}\,(L_L)_{\beta d},\nonumber \\
&&O^{(2)}_{abcd}=\epsilon_{ijk}\epsilon_{\alpha\beta} 
\overline{(Q_L)^c_{i\alpha a}}(Q_{L})_{j\beta b}
\overline{(u_R)^c_{kc}}\,(l_{R})_{d},\nonumber \\
&&O^{(3)}_{abcd}=\epsilon_{ijk}\epsilon_{\alpha\beta}\epsilon_{\gamma\rho}
\overline{(Q_L)^c_{i\alpha a}}(Q_L)_{j\beta b}\overline{(Q_L)^c_{k\gamma c}}
\,(L_L)_{{\rho d}},
\nonumber \\
&&O^{(4)}_{abcd}=\epsilon_{ijk}(\vec{\tau}\epsilon)_{\alpha\beta}\cdot
(\vec{\tau}\epsilon)_{\gamma \rho}
\overline{(Q_L)^c_{i\alpha a}}(Q_L)_{j\beta b}\overline{(Q_L)^c_{k\gamma c}}
\,(L_L)_{{\rho d}},
\nonumber \\
&&O^{(5)}_{abcd}=\epsilon_{ijk}\overline{(d_R)^c_{ia}}(u_R)_{jb}
\overline{(u_R)^c_{kc}}\,(l_{R})_{d},\nonumber \\
&&O^{(6)}_{abcd}=\epsilon_{ijk}\overline{(u_R)^c_{ia}}(u_R)_{jb}
\overline{(d_R)^c_{kc}}\,(l_{R})_{d},\nonumber \\
&&O^{(7)}_{abcd}=\epsilon_{ijk}\overline{(u_R)^c_{ia}}(d_R)_{jb}
\overline{(d_R)^c_{kc}}\,(\nu_{R})_{d},\nonumber \\
&&O^{(8)}_{abcd}=\epsilon_{ijk}\epsilon_{\alpha\beta}
\overline{(d_R)^c_{ia}}(Q_L)_{j\alpha b}
\overline{(\nu_R)_{c}}\,(Q_L)_{k\beta d},
\label{effective}
\end{eqnarray} 
where $i,j,k$ are $SU(3)$ indices; $\alpha,\beta,\gamma$ and $\rho$ are $SU(2)$
indices; and $a,b,c$ and $d$ are generation indices. From Eq.~(\ref{effective}),
using Fierz transformations, it is possible to obtain all vector and tensor
Dirac matrices~\cite{sw79}. All operators in Eq.~(\ref{effective}) are
forbidden by the $Z_{13}$ symmetry in Eq.~(\ref{z13}); $d=7$ operators formed
with those of Eq.~(\ref{effective}) and the singlet $\phi$ (or $\phi^*$) are
also forbidden. Notice that $O^{(2,5,6,7,8)}_{abcd}$ are also forbidden by
$Z_3$. However, there are others $B-L$ conserving operators allowed by all the 
symmetries of the model as
\begin{equation}
O^{(1)}_{abcd}\,\Phi^\dagger_l\Phi_u,\;
O^{(1)}_{abcd}\,\phi^4, \; 
O^{(3,4)}_{abcd}\,\phi^5, 
\label{more}
\end{equation}
of $d=8,10,11$, respectively, that may induce the proton decay, via four
fermion interactions, after the spontaneous symmetry breaking. Let us write the
proton lifetime as  
\begin{equation}
\tau_P\propto 
\tau^{5}_P\,\left(\frac{\alpha_5}{\alpha_U}\right)\;
\left(\frac{M_U}{M_5}\right)^4\,\vert \xi\vert^{-2}
\label{plifetime}
\end{equation} 
where $\tau^5_P=M^4_5\alpha^{-1}_5 m^{-5}_P$ with 
$M_5$ is the unification scale in the context of the MSSM, $M_5\simeq 2.1\times
10^{16}$ GeV; $\alpha_5$ is the respective coupling constant at that 
unification scale with $\alpha^{-1}_5=24$; $m_P$ is the proton mass; and
$\alpha_U$ is the coupling constant at $M_U$ in this model with
$\alpha^{-1}_U=38$; finally, $\xi$ is a factor depending on the effective
operator. Although the $d=8$ operator is suppressed by $1/M^{4}_U$, 
after the spontaneous symmetry breaking, it induces a four fermion 
interaction proportional to $\xi=v^*_lv_u/M^2_U$. Since $\vert
v_lv_u\vert\lesssim (246\, {\rm GeV})^2$ we have that 
$\vert\xi\vert<7.7\times 10^{-23}$ and since $M_U/M_5\simeq 
1.3\times10^{-3}$, $\alpha_5/\alpha_U = 38/24\simeq 1.6$,
in Eq.~(\ref{plifetime}) there is a factor $\stackrel{>}{\sim}8\times
10^{32}$ with respect to $\tau^5_P$. The $d=10$
operator is suppressed by $M^{-2}_U\Lambda^{-4}$ and it induces four 
fermion interactions like $M^{-2}_U(v_\phi/\Lambda)^4 O^{(1)}$ where $\Lambda$
is a mass scale connecting the field $\phi$ with the four fermion effective
operators $O^{(i)}$, $\Lambda$ may be $M_U$ (or the PQ scale) 
or the Planck scale. In this case there is
a factor $\vert\xi\vert^{-2} =(\Lambda/v_\phi)^8$ in Eq.~(\ref{plifetime}). 
The enhancement on the proton lifetime depends on the scales $\Lambda$ and
$v_\phi$. Assuming $v_\phi=10^{12}$ GeV and $\Lambda=M_{\rm Planck}=10^{19}$
GeV, we have an enhancement factor of $5\times 10^{44}$ with respect
$\tau^5_P$. If, instead of $M_{\rm Plank}$ we use $\Lambda=M_U$ but
$v_\phi=10^9$ GeV we still obtain an enhancement factor $1.7\times 10^{24}$ in
the proton lifetime. Finally, if $v_\phi=10^{12}$ GeV and $\Lambda=M_U$ the
proton lifetime is raised by a factor two with respect to
$\tau^5_P$. Similar analysis follows for the $d=11$ effective operators. Hence, 
this model survive the proton decay problem since 
with the natural values of the parameters we have that the proton
has a lifetime which is compatible with the no observation of its decays at the
present experimental level. Moreover, notice that the $d=5$ effective operator
$M_U^{-1}LL\Phi_\nu\Phi_\nu$ is allowed by the $Z_{13}$ symmetry but forbidden  
by $Z_3$. However the $d=10$ operator
$M_U^{-1}\Lambda^{-5}LL\Phi_\nu\Phi_\nu\phi^5$ gives a Majorana mass to the
neutrinos with a upper limit of 2 eV, obtained when $\Lambda=v_\phi$ and 
$\langle\Phi_\nu\rangle=246$ GeV. 
 
Summarizing, we have obtained a multi-Higgs extension of the standard model
with $Z_{13}\otimes Z_3$ symmetries that imply an automatic PQ, $B$ and $L$ 
symmetries at the tree level. The axion is stabilized against semiclassical
gravitational effects by those symmetries and they also stabilize the nucleon
allowing, at the same time, the unification of the three gauge coupling
constants at an energy near the PQ scale. Last but not least, the correct value
of the weak mixing angle at the $Z$-peak is obtained. Although we can always
implement a larger $Z_N$ by adding more matter multiplets, concerning the
unification of the coupling constants, a larger number of multiplets or higher
dimensional representation of $SU(2)$ affect the running of the couplings. Only
a limited set of representations is allowed in this respect.
We should mention that an unification scale near the PQ scale is also obtained
in an $[SU(3)]^4$ model~\cite{bmw} but this model has no PQ symmetry in its
minimal version. The present model cannot be
supersymmetric at low energy (of the order of TeVs), since the fermion
superpartners of the Higgs scalars would upset the unification of 
the gauge couplings, however it is possible to have supersymemtry if
$\mu_{SUSY}\stackrel{>}{\sim}M_U$. It would be interesting to search what sort
of non-SUSY GUT embed this model.     

\acknowledgments 
A. G. D. was supported by FAPESP under the process 01/13607-3, and V. P. was
supported partially by CNPq under the process 306087/88-0. A. D. G. would also
thanks to J. K. Mizukoshi for helping with the figures.


\begin{thebibliography}{99}

\bibitem{gg74} H. Georgi and S. L. Glashow, Phys. Rev. Lett. {\bf32}, 438
(1974).
\bibitem{hquinn} H. Georgi, H. Quinn, and S. Weinberg, Phys. Rev. Lett. {\bf33},
451 (1974).
\bibitem{gut} P. Langacker, Phys. Rep. {\bf72}, 185 (1981).
\bibitem{gut2} R. Slansky, Phys. Rep. {\bf 79}, 1 (1981).
\bibitem{pdg} K. Hagiwara \textit{et al.}, (Particle Data Group), Phys. Rev. D
{\bf66}, 010001 (2002). 
\bibitem{susy} U. Amaldi \textit{et al.}, Phys. Rev. D {\bf36}, 1385 (1987); 
P. Langacker and M. Luo, Phys. Rev. D {\bf44}, 817 (1991).
\bibitem{sk99} Y. Hayato \textit{et al.}, (SuperKamiokande Collaboration), 
Phys. Rev. Lett. {\bf83}, 1529 (1999).
\bibitem{d5} S. Weinberg, Phys. Rev. D {\bf26}, 287 (1982); 
P. Nath, A. H. Shamseddine, and R. Arnowitt, Phys. Rev. D {\bf32}, 2348 (1985);
T. Goto and T. Nihei, Phys. Rev. D {\bf59}, 115009 (1999); H. 
Murayama and A. Pierce, Phys. Rev. D {\bf65}, 055009 (2002); B. Bajc, P. F.
Perez, and G. Senjanovic, Phys. Rev. D {\bf66},   
075005 (2002), and references therein.
\bibitem{perez}  P. F. Perez, hep-ph/0403286.
\bibitem{willenbrock} S. Willenbrock, Phys. Lett. {\bf B561}, 130 (2003). 
\bibitem{berezhiani01} Z. Berezhiani, I. Gogoladze, and A. Kobakhidze, Phys.
Lett. {\bf B522}, 107 (2001).   
\bibitem{bmw} K. S. Babu, E. Ma, and S. Willenbrock, Phys. Rev. D {\bf 69}, 
051301 (2004). 
\bibitem{dine} M. Dine, W. Fischler, and M. Srednicki, Phys. Lett. {\bf104B},
199 (1981).
\bibitem{iaxionsm} A. G. Dias, V. Pleitez, and M. D. Tonasse, Phys. Rev. D
{\bf69}, 015007 (2004).
\bibitem{gogoladze} K. S. Babu, I. Gogoladze, and K. Wang, Phys. Lett. {\bf
B570}, 32 (2003); I. Gogoladze, hep-ph/0402087.
\bibitem{outros} 
A. G. Dias, V. Pleitez, and M. D. Tonasse, Phys. Rev. D {\bf67}, 095008 (2003); 
K. S. Babu, I. Gogoladze, and K. Wang, Nucl. Phys. {\bf B 660}, 322 (2003); 
K. S. Babu, I. Gogoladze, and K. Wang, Phys. Lett. {\bf B 560}, 214 (2003); 
A. G. Dias and V. Pleitez, Phys. Rev. D{\bf 69}, 077702 (2004); A. G. Dias, C.
A.  de S. Pires, and P. S. Rodrigues da Silva, Phys. Rev. D {\bf68}, 115009
(2003).
\bibitem{marciano82} W. Marciano and G. Senjanovic, Phys. Rev. D {\bf25}, 3092
(1982).
\bibitem{jones} D. R. T. Jones, Phys. Rev. D{\bf 25}, 581 (1982); M. E. Machacek
and M. T. Vaugh, Nucl. Phys. {\bf B222}, 83 (1983).
\bibitem{pq} R. D. Peccei and H. Quinn, Phys. Rev. Lett. {\bf38}, 1440 (1977).
\bibitem{sw79} S. Weinberg, Phys. Rev. Lett. {\bf43}, 1566 (1979). 

\end{thebibliography}
\end{document}